\PassOptionsToPackage{hyperfootnotes=false}{hyperref}
\documentclass[a4paper,fleqn]{cas-dc}

\usepackage[numbers,sort&compress]{natbib}
\usepackage{makecell}
\graphicspath{{figures/}}

\newcolumntype{P}[1]{>{\raggedright\arraybackslash}p{#1}}

\begin{document}

\let\WriteBookmarks\relax

\shorttitle{Exact Lyapunov spectra of affine cellular automata}
\shortauthors{M.~Rollier and J.~M.~Baetens}

\title[mode=title]{Exact Lyapunov spectra of affine cellular automata and the parity rule on networks}

\author[ugent,usp]{Michiel Rollier}[orcid=0000-0001-8467-734X]
\cormark[1]
\ead{michiel.rollier@ugent.be}
\credit{Conceptualization, Methodology, Software, Formal analysis, Validation,
Visualization, Writing -- original draft}

\author[ugent]{Jan M.~Baetens}[orcid=0000-0003-4084-9992]
\credit{Funding acquisition, Supervision, Writing -- review \& editing}

\affiliation[ugent]{organization={BionamiX, Dept.~of Data Analysis and
Mathematical Modelling, Ghent University},
            addressline={Coupure Links, 653},
            city={Ghent},
            postcode={9000},
            country={Belgium}}

\affiliation[usp]{organization={S\~{a}o Carlos Institute of Physics, University of S\~{a}o Paulo},
            addressline={Av.~Trab.~S\~{a}o Carlense, 400},
            city={S\~{a}o Carlos},
            postcode={13566-590},
            country={Brazil}}

\cortext[cor1]{Corresponding author.}

\begin{abstract}
The Lyapunov exponent quantifies the sensitivity of a dynamical system to perturbations, and the full Lyapunov spectrum extends this to every orthogonal direction in tangent space. For cellular automata the spectrum is almost always approximated numerically, and the approximation is delicate. We show that the affine rules, those whose update is an \textsc{xor} of a subset of the inputs together with a constant, admit an exact Lyapunov spectrum. An affine rule has a configuration-independent Boolean Jacobian, so the spectrum reduces to the logarithms of the eigenvalue moduli of a single constant matrix, with no simulation and no limit involved. Two cases are shown to carry a closed form. For an affine cellular automaton on a periodic lattice the Jacobian is a multilevel circulant matrix, and its eigenvalues are given by the discrete Fourier transform of the rule's gradient stencil, valid in any spatial dimension. For the parity rule on an arbitrary graph the Jacobian is the adjacency matrix, plus the identity in the self-inclusive case, so the spectrum is the logarithm of the absolute spectrum of that matrix, and the maximal exponent the logarithm of its spectral radius. In both cases the exact spectra serve as benchmarks: they expose errors in previously reported spectra, and turn the informal correspondence between spectral radius and tangent-space sensitivity into an exact identity.
\end{abstract}


\maketitle

% ============================================================
\section{Introduction}
\label{sec:intro}

The Lyapunov exponent measures how fast nearby trajectories of a dynamical system separate; a positive exponent is the signature of exponential sensitivity and is often deployed as the definition of chaotic behaviour. For continuous systems the notion is classical, and the full spectrum of exponents, one per orthogonal direction in tangent space, is a standard descriptor of stability \cite{eckmann1985ergodic}. For cellular automata (CAs) the situation is less settled, because the state space is discrete and a derivative of the local update rule is not available in the ordinary sense \cite{vispoel2022progress}. Two routes have been taken. The symbolic-dynamics route defines left and right exponents through the shift map on bi-infinite configurations and relates them to entropy \cite{shereshevsky1992lyapunov,tisseur2000cellular}. The damage-spreading route, which we follow here, replaces the derivative by the Boolean derivative of Vichniac \cite{vichniac1990boolean} and tracks the growth of a perturbation through the resulting Boolean Jacobian \cite{bagnoli1992damage,bagnoli1999synchronization}. This second route leads naturally to Lyapunov profiles for CAs \cite{baetens2016stability,baetens2018lyapunov} and to a Lyapunov spectrum for CAs and for network automata (NAs) \cite{vispoel2024damage}.

Computing such a spectrum is, however, numerically demanding. The spectrum is read from the singular values of a long product of Jacobians, and calculating these Jacobians generally requires the exact evolution of the state vectors over subsequent timesteps. Making matters worse, the ratio between the largest and the smallest of these singular values grows exponentially with the number of timesteps. Direct multiplication therefore loses the slower-growing directions to floating-point error, and a stable approximation requires repeated re-orthonormalisation, as in the algorithm of Benettin et al.~\cite{benettin1980lyapunov1,benettin1980lyapunov2}.

In this paper we show that a particular subclass of Boolean synchronous deterministic local update rules escapes these difficulties altogether. This subclass consists of rules whose Jacobian does not depend on the configuration, but is simply a single constant matrix. This implies that one does not need to calculate a trajectory or take a limit, because the product of Jacobians collapses to a matrix power, and the Lyapunov spectrum is exactly the set of logarithms of the moduli of that matrix's eigenvalues.

We will consider rules that live on CAs with periodic lattices (starting with elementary CAs), and rules that live on NAs on arbitrary undirected unweighted simple graphs. For the CAs, rules with a constant Jacobian are exactly the \emph{affine} rules. A rule is affine when its Boolean update function is an \textsc{xor} of a subset of its inputs together with a Boolean constant. In this case, the Jacobian is a (multilevel) circulant matrix: its eigenvalues are given by the discrete Fourier transform (DFT) of the rule's gradient stencil, and its singular values are their moduli. This yields a closed-form Lyapunov spectrum for affine CAs in any dimension and for any neighbourhood that is the same at every cell. The notion of an affine rule on CAs does not carry over cleanly to an arbitrary graph, because then a local update rule can only be uniform if it is (outer)totalistic. For that reason, the only affine NA rule that is both neighbour-dependent and well defined on every graph is the parity rule: the \textsc{xor} of the entire neighbourhood, with or without the node itself. For the parity rule the constant Jacobian is the adjacency matrix itself, shifted by the identity in the self-inclusive case, which yields exact relationships between the parity rule's Lyapunov spectrum and the graph spectrum.

Linear and affine rules have, of course, been studied at length \cite{hudcova2024simulation}. The global dynamics of linear rules over $\mathrm{GF}(2)$ are classical \cite{martin1984algebraic,manzini1999attractors}, and closed-form Lyapunov exponents for linear CAs are already available in the symbolic-dynamics route, where they record the left and right speeds at which a difference spreads through a bi-infinite configuration \cite{damico2003computing}. That is a different object from the damage-spreading spectrum used here, which is read from the singular values of the product of Boolean Jacobians on a finite lattice or graph \cite{vispoel2024damage}. Our contribution is that, in this damage-spreading setting, an affine rule has a constant Jacobian. On periodic lattices and undirected graphs this Jacobian is also normal, so the full spectrum follows exactly and in closed form from its eigenvalues, rather than being approximated.

The paper is organised as follows. Sec.~\ref{sec:jacobian} sets up the Boolean Jacobian and the Lyapunov spectrum, and distinguishes the configuration- and tangent-space pictures. Sec.~\ref{sec:lattice} outlines the relationship between affine CA rules and constant Jacobians in more detail, and works out the lattice case and its closed form. Sec.~\ref{sec:parity-network} treats the parity rule on networks, linking the Lyapunov spectrum and the graph spectrum. Sec.~\ref{sec:conclusion} concludes and briefly considers next steps. We list all technical abbreviations in Tab.~\ref{tab:abbreviations} for the reader's convenience.

\begin{table}[t]
    \centering
    \caption{Abbreviations used in this article.}
    \label{tab:abbreviations}
    \begin{tabular*}{\tblwidth}{@{} l l @{}}
    \toprule
    \textbf{Abbreviation} & \textbf{Meaning}\\
    \midrule
    CA      & cellular automaton\\
    DDS     & discrete dynamical system\\
    DFT     & discrete Fourier transform\\
    ECA     & elementary cellular automaton\\
    $\mathrm{GF}(2)$ & Galois field of two elements\\
    MLE     & maximal Lyapunov exponent\\
    $n$-D   & $n$-dimensional\\
    NA      & network automaton\\
    v.~N.  & von Neumann (neighbourhood)\\
    \bottomrule
    \end{tabular*}
\end{table}

% ============================================================
\section{The Boolean Jacobian and the Lyapunov spectrum}
\label{sec:jacobian}

\subsection{The Boolean derivative and Jacobian}

We consider a binary discrete dynamical system (DDS) with state set $\mathcal{S}=\{0,1\}$. In this article we consider two kinds of DDS, namely a CA with $N$ cells $\{c_i\}_{i=1}^N$ on a regular lattice, and an NA with $N$ nodes $\{v_i\}_{i=1}^N$ on an arbitrary graph. The dynamics on the DDS are governed by a local update rule $\phi$ that induces a global update function $\Phi:\mathcal{S}^N\to\mathcal{S}^N$. A configuration is a vector $\mathbf{s}=(s_1,\dots,s_N)$ with $s_i\in\mathcal{S}$, and the configuration at timestep $t$ is written $\mathbf{s}^t$. We refer to \cite{rollier2025comprehensive} for a taxonomy of these systems and the relations between them.

The Jacobian generalises the derivative to a vector-valued function. For a binary DDS its entries are given by the Boolean derivative $J_{ij}(\mathbf{s})=\partial\Phi_i/\partial s_j$ \cite{vichniac1990boolean,bagnoli1992damage}, where
\begin{equation}
    \label{eq:boolean-derivative}
    \frac{\partial\Phi_i}{\partial s_j}
    =\Phi(s_1,\dots,s_j,\dots,s_N)_i\;\oplus\;
     \Phi(s_1,\dots,\bar{s}_j,\dots,s_N)_i,
\end{equation}
$\oplus$ is the \textsc{xor} operator and $\bar{s}_j=1\oplus s_j$. The entry $J_{ij}$ answers a simple question: if the state of cell $c_j$ (or node $v_j$) is flipped, does the state of cell $c_i$ (or node $v_i$) change at the next timestep? If it does, $J_{ij}=1$; if it does not, $J_{ij}=0$. The derivative is evaluated by flipping $s_j$, applying $\Phi$ to both configurations, and comparing the $i$-th output; carrying this out for every pair $(i,j)$ assembles the whole matrix. By construction the Jacobian is Boolean, $\mathbf{J}\in\{0,1\}^{N\times N}$, and it is contained in the adjacency matrix $\mathbf{A}$ of the underlying topology, in the sense that $A_{ij}=0$ implies $J_{ij}=0$ for $i\neq j$: a cell cannot influence a cell to which it is not connected. A rule may, however, depend on a cell's own state, so a diagonal entry $J_{ii}$ may be nonzero even when $A_{ii}=0$; allowing for this, $\mathbf{J}\subseteq\mathbf{A}+\mathbf{I}$. In general, $\mathbf{J}$ depends on the configuration $\mathbf{s}$, and we write
$\mathbf{J}(\mathbf{s}^t)$ for the Jacobian at timestep $t$.

\subsection{The Jacobian of elementary CAs}

For an elementary cellular automaton (ECA), the local rule $\phi$ acts on a radius-one neighbourhood, mapping the states of a triplet of adjacent cells $(c_-, c_\circ, c_+)$ at timestep $t$ to the state of the triplet's central cell $c_\circ$ at the next timestep. This rule can be reduced to a Boolean expression: Wolfram rule $30$, for example, is expressed as $(s_-, s_\circ, s_+) \mapsto s_- \oplus (s_\circ + s_+)$, where $+$ denotes the \textsc{or} operator. The \emph{gradient} of $\phi$ evaluated at some state triplet $(s_{i-1}, s_i, s_{i+1})$ in $\mathbf{s}$ is then 
\begin{equation}
    \label{eq:eca-gradient}
    \nabla\phi(s_{i-1}, s_i, s_{i+1}) = (J_{i,i-1},J_{i,i},J_{i,i+1}).
\end{equation}
Each component is the Boolean derivative \eqref{eq:boolean-derivative} of the local rule with respect to the corresponding neighbour, reduced to a Boolean expression. The gradient of every ECA can be read off from the local rule in this way. Recomputing it in disjunctive normal form for the $88$ non-equivalent ECAs reproduces the table of Vichniac \cite{vichniac1990boolean} except for five rules ($62$, $110$, $130$, $146$ and $172$), whose entries are misprinted in the original publication. The corrected entries are given in Tab.~\ref{tab:gradient-table} and we show their derivation in App.~\ref{app:vichniac}. These gradients allow one to construct the complete Jacobian, which is then used to track perturbations in tangent space.

\begin{table}[t]
    \centering
    \caption{Corrected gradients in disjunctive normal form for the five ECAs whose published
    values in \cite{vichniac1990boolean} are misprints. Here juxtaposition denotes \textrm{\textsc{and}}, $+$ denotes \textrm{\textsc{or}}, and an overbar denotes \textrm{\textsc{not}}.}
    \label{tab:gradient-table}
    \begin{tabular*}{\tblwidth}{@{}r l l @{}}
    \toprule
    \textbf{Rule} & $\phi(s_{i-1},s_i,s_{i+1})$ & $\nabla\phi(s_{i-1},s_i,s_{i+1})$\\
    \midrule
    62 & $s_{i-1}\bar{s}_i+s_i\bar{s}_{i-1}+s_{i+1}\bar{s}_{i-1}$ & $(s_i+\bar{s}_{i+1},s_{i-1}+\bar{s}_{i+1},\bar{s}_{i-1}\bar{s}_i)$\\
    110 & $s_i\bar{s}_{i+1}+s_{i+1}\bar{s}_{i-1}+s_{i+1}\bar{s}_i$ & $(s_is_{i+1},s_{i-1}+\bar{s}_{i+1},s_{i-1}+\bar{s}_i)$ \\
    130 & $s_{i-1}s_is_{i+1}+s_{i+1}\bar{s}_{i-1}\bar{s}_i$ & $(s_{i+1},s_{i+1},s_{i-1}s_i+\bar{s}_{i-1}\bar{s}_i)$ \\
    146 & \makecell[tl]{$s_{i-1}s_is_{i+1}+s_{i-1}\bar{s}_i\bar{s}_{i+1}$ \\ $\ +\,s_{i+1}\bar{s}_{i-1}\bar{s}_i$} & $(s_{i+1}+\bar{s}_i,s_{i-1}+s_{i+1},s_{i-1}+\bar{s}_i)$\\
    172 & $s_{i-1}s_{i+1}+s_i\bar{s}_{i-1}$ & $(s_i\bar{s}_{i+1}+s_{i+1}\bar{s}_i,\bar{s}_{i-1},s_{i-1})$\\
    \bottomrule
    \end{tabular*}
\end{table}

\subsection{Configuration space and tangent space}
\label{subsec:two-spaces}

A perturbation admits two distinct descriptions, and the distinction governs what follows; we adopt the terminology of Vispoel et al.~\cite{vispoel2024damage}.

\begin{figure}
    \centering
    \includegraphics[width=\linewidth]{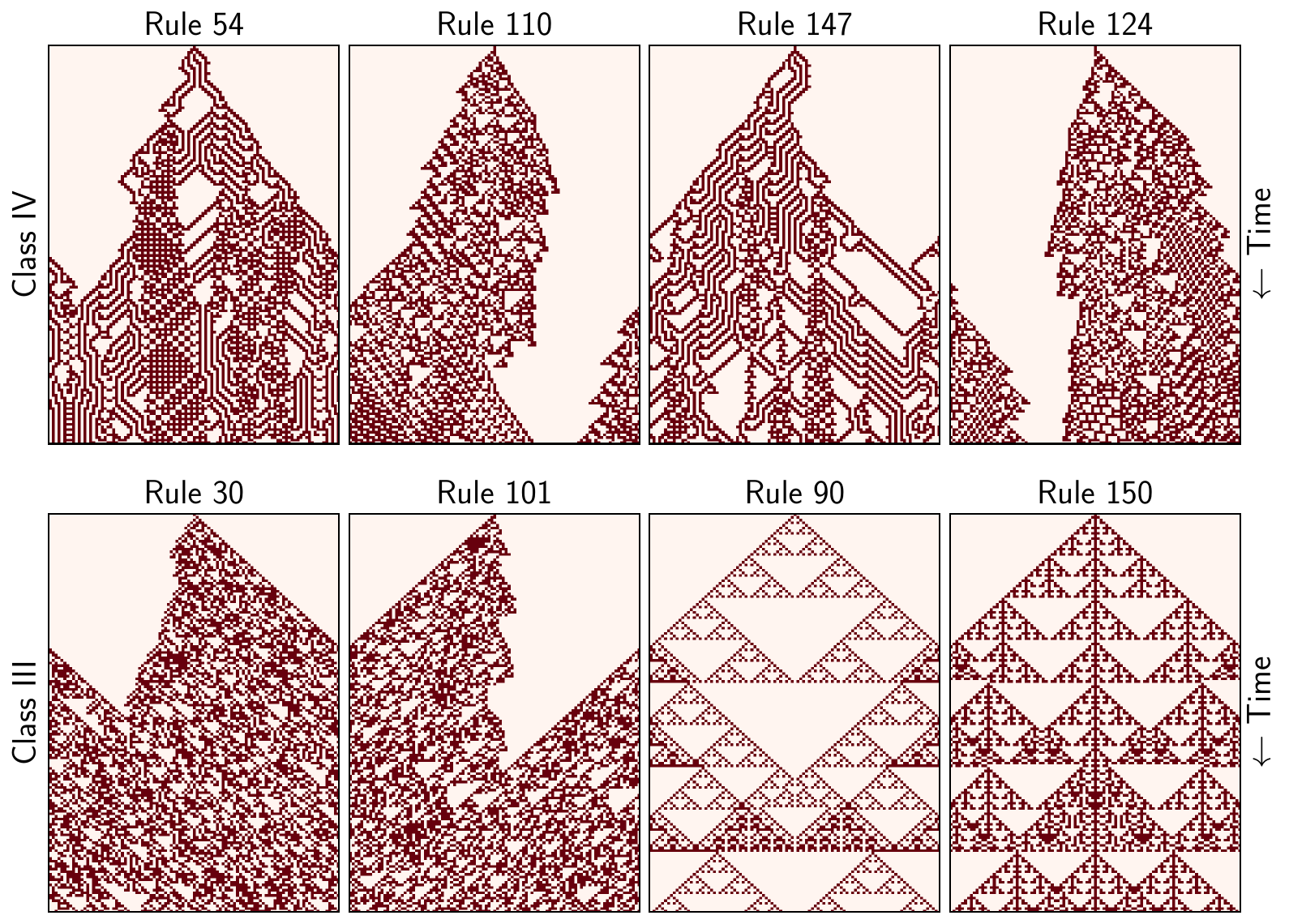}
    \caption{Configuration-space difference patterns for representative ECA rules from a random initial configuration. A perturbation is introduced in a single cell and the mismatching cells are recorded over time.}
    \label{fig:defect-cones}
\end{figure}

Let two configurations differ in a single cell $c_j$, such that $\Delta\mathbf{s}^0=\mathbf{e}_j$. The \emph{difference pattern} $\Delta\mathbf{s}^t=\mathbf{s}^t\oplus\hat{\mathbf{s}}^t$, with $\hat{\mathbf{s}}^t$ evolved from $\mathbf{s}^0\oplus\Delta\mathbf{s}^0$, records which cells actually differ. The Jacobian reproduces it exactly for one step,
\begin{equation}
    \label{eq:one-step}
    \Delta\mathbf{s}^1=\mathbf{J}(\mathbf{s}^0)\,\Delta\mathbf{s}^0 ,
\end{equation}
but the relation does not iterate: the defect then occupies several cells, and the response of a Boolean rule to simultaneous flips is not the \textsc{xor} of its responses to each separately, so in general $\Delta\mathbf{s}^t\neq\mathbf{J}(\mathbf{s}^{t-1})\,\Delta\mathbf{s}^{t-1}$ for $t>1$. This is not the controlled error of a linearisation: the smallest perturbation is a whole cell, and no approximation theorem bounds the discrepancy. Since information travels at most one cell per timestep in each direction, the pattern remains confined to an expanding cone (Fig.~\ref{fig:defect-cones}).

Applying the Jacobians in sequence along a single trajectory gives instead
\begin{equation}
    \label{eq:jacobian-product}
    \mathbf{J}^{(t)}=\mathbf{J}(\mathbf{s}^{t-1})\,\mathbf{J}(\mathbf{s}^{t-2})
    \cdots\mathbf{J}(\mathbf{s}^0),
    \qquad \delta\mathbf{s}^t=\mathbf{J}^{(t)}\,\delta\mathbf{s}^0 ,
\end{equation}
with $\delta\mathbf{s}^0=\Delta\mathbf{s}^0$; we reserve $\Delta$ for the Boolean difference pattern and $\delta$ for the tangent-space perturbation. These products are taken over the integers, which is what separates the two descriptions. Since $J_{ij}=1$ records that $c_j$ influences $c_i$ in one step, $(\mathbf{J}^{(t)})_{ij}$ counts the walks of influence from $c_j$ to $c_i$ in $t$ steps. The tangent-space perturbation counts these walks. The difference pattern does not: two walks arriving at a cell reinforce one another over $\mathbb{Z}$ but cancel over $\mathrm{GF}(2)$. For an affine rule the two objects are thus not a quantity and its approximation but one matrix in two arithmetics, and the same binary matrix can yield an unbounded integer count alongside a bounded Boolean pattern. Empirically, perturbations in both spaces are loosely related, as we show in App.~\ref{app:mle-vs-damage}.

If tangent-space dynamics do not describe the difference patterns of actual trajectories, one may reasonably ask what has earned them their place. The first answer is that exponential divergence can be defined nowhere else. A finite system such as a CA with periodic boundaries eventually cycles, so the difference pattern between any two trajectories is bounded and eventually periodic and its growth rate is zero for every rule; chaos in the strict sense is impossible in such systems, as Luque and Sol\'e~\cite{luque2000lyapunov} note. Only the walk count can grow exponentially. The second answer is that tangent space is where the classical stability theory of Boolean networks is built. While a defect still occupies few cells, walks rarely meet, so the walk count and the difference pattern agree; on annealed or locally tree-like networks this holds for some time, whereas on lattices and undirected graphs walks can already meet at $t=2$. Since Derrida and Pomeau~\cite{derrida1986random}, the order-chaos transition has been located in this regime, by asking whether a small perturbation grows or decays on average rather than by following a particular pair of trajectories. Luque and Sol\'e~\cite{luque2000lyapunov} obtain the corresponding exponent from a mean-field treatment of the Boolean Jacobian, and Pomerance et al.~\cite{pomerance2009effect} show, in an approximation for large networks of any topology, that stability is set by the largest eigenvalue of an averaged Jacobian. For the parity rule every flip propagates, so this averaged Jacobian is $\mathbf{A}+a_\circ\mathbf{I}$ itself and the stability criterion coincides with the sign of the MLE (Sec.~\ref{sec:parity-network}). The tangent-space exponent is therefore the quantity that this stability theory uses.

\subsection{The Lyapunov spectrum}
\label{subsec:lyapunov-spectrum}

The tangent-space object we will focus on here is the Lyapunov spectrum, recording the growth rate along every orthogonal direction in tangent space simultaneously. Following the established treatment for continuous-state systems \cite{eckmann1985ergodic}, we track the full ellipsoid of perturbations and record the growth of its semi-axes. These semi-axes are the singular values $\{\sigma_k\}_{k=1}^N$ of $\mathbf{J}^{(t)}$, and the Lyapunov spectrum collects their growth rates,
\begin{equation}
    \label{eq:lyapunov-spectrum}
    \Lambda_k=\lim_{t\to\infty}\frac{1}{t}\,\ln\bigl(\sigma_k\bigr),
    \qquad k=1,\dots,N.
\end{equation}
The largest element is the maximal Lyapunov exponent (MLE), whose sign indicates whether tangent-space perturbations grow or decay \cite{eckmann1985ergodic}.

Working with the singular values is a deliberate choice. The singular values are invariant under orthogonal changes of basis, so the spectrum is a property of the rule and not of the cell labelling; for the normal constant-Jacobian rules studied below it is the logarithm of the moduli of the eigenvalues of $\mathbf{J}$, which is exactly what connects it to the algebraic structure of the rule. The orthogonal directions also separate growth rates that a cell-by-cell view conflates, since a generic perturbation in tangent space aligns with the fastest-growing direction and grows at the maximal rate. The sum of all exponents, finally, measures volume change in tangent space, and the sum of the positive exponents links the spectrum to the Kolmogorov-Sinai entropy, a connection examined empirically for the damage-spreading spectrum in \cite{vispoel2024damage}.

Three obstacles stand in the way of computing Eq.~\eqref{eq:lyapunov-spectrum} in general. First, the limit must be replaced by a finite cut-off. Second, the Jacobians depend on the configuration, so the spectrum must be averaged over an ensemble of initial configurations. Third, the computation is numerically fragile: as $t$ grows the ratio between the largest and smallest singular values of $\mathbf{J}^{(t)}$ quickly exceeds floating-point precision, so the product cannot be formed directly. We show next that all three obstacles disappear at once for rules with a constant Jacobian, for which the Lyapunov spectrum is an analytical quantity, and which for that reason serve as benchmarks.

% ============================================================
\section{Affine cellular automata}
\label{sec:lattice}

\subsection{Affine ECAs have a constant Jacobian}
\label{sec:affine}

A binary rule is \emph{affine} when its local update function is an \textsc{xor} of a subset of the neighbourhood states and a binary constant. For an ECA this means
\begin{equation}
    \label{eq:affine-eca}
    \phi(s_{i-1},s_i,s_{i+1})
    =a_0\,\oplus\,a_-\,s_{i-1}\,\oplus\,a_\circ\,s_i\,\oplus\,a_+\,s_{i+1},
\end{equation}
and the affine rules are the eight linear rules ($a_0=0$) together with their complements ($a_0 = 1$) (Tab.~\ref{tab:affine-ecas}). Their global dynamics are classical: they reduce to linear algebra over the field $\mathrm{GF}(2)$, which is why the state-transition structure of linear rules has been characterised in detail \cite{martin1984algebraic}, as has their asymptotic behaviour on infinite lattices \cite{manzini1999attractors}. Our concern is not the global dynamics but the Lyapunov spectrum, and for that the relevant property is the following.

\begin{table}[t]
\centering
\caption{The $16$ affine ECAs (nonlinear ECAs in parentheses), grouped by the number of inputs the rule is sensitive to (the gradient weight $\lvert\nabla\phi\rvert$). The constant term $a_0$ does not enter the Jacobian, so each rule and its complement share a gradient.}
\label{tab:affine-ecas}
\begin{tabular*}{\tblwidth}{@{} p{2cm} p{3.2cm} c c @{}}
\toprule
\textbf{Rule} & $\phi$ & $\nabla\phi$ & $\lvert\nabla\phi\rvert$\\
\midrule
0 (255)   & $0\,(\oplus 1)$                                 & $(0,0,0)$ & 0\\
170 (85)  & $s_{i+1}\,(\oplus 1)$                           & $(0,0,1)$ & 1\\
204 (51)  & $s_i\,(\oplus 1)$                               & $(0,1,0)$ & 1\\
240 (15)  & $s_{i-1}\,(\oplus 1)$                           & $(1,0,0)$ & 1\\
60 (195)  & $s_{i-1}\oplus s_i\,(\oplus 1)$                 & $(1,1,0)$ & 2\\
90 (165)  & $s_{i-1}\oplus s_{i+1}\,(\oplus 1)$             & $(1,0,1)$ & 2\\
102 (153) & $s_i\oplus s_{i+1}\,(\oplus 1)$                 & $(0,1,1)$ & 2\\
150 (105) & $s_{i-1}\oplus s_i\oplus s_{i+1}\,(\oplus 1)$   & $(1,1,1)$ & 3\\
\bottomrule
\end{tabular*}
\end{table}

For an affine rule the Boolean derivative is trivial to evaluate. Flipping the state of the left neighbour $s_{i-1}$ in Eq.~\eqref{eq:affine-eca} toggles the output precisely when $a_-=1$, independently of the rest of the configuration, and likewise for $s_i$ and $s_{i+1}$. Each Boolean derivative is therefore the constant coefficient of the corresponding input, so $\nabla\phi=(a_-,a_\circ,a_+)$ and the Jacobian does not depend on $\mathbf{s}$. The converse also holds: if every Boolean derivative is a constant then the update is \textsc{xor}-linear in each argument, hence affine. So a rule is affine if and only if its Boolean Jacobian is configuration-independent.

When the Jacobian is constant, the product \eqref{eq:jacobian-product} collapses to the matrix power $\mathbf{J}^{(t)}=\mathbf{J}^t$. The constant Jacobians we encounter are normal (a circulant matrix on a periodic lattice, Eq.~\eqref{eq:circulant-J}, and a real symmetric matrix on an undirected graph, Sec.~\ref{sec:parity-network}), hence unitarily diagonalisable, so the singular values of $\mathbf{J}^t$ are the moduli $\lvert\lambda_k(\mathbf{J})\rvert^t$. Substituting into Eq.~\eqref{eq:lyapunov-spectrum}, the factor $t$ cancels, the limit becomes an equality, and
\begin{equation}
    \label{eq:exact-spectrum}
    \Lambda_k=\ln\lvert\lambda_k(\mathbf{J})\rvert=\ln\bigl(\sigma_k(\mathbf{J})\bigr),\qquad k=1,\dots,N,
\end{equation}
the two forms agreeing because a normal matrix has singular values equal to its
eigenvalue moduli. Normality is essential here: for a non-normal constant
Jacobian the singular values of $\mathbf{J}^t$ are not the powers of those of
$\mathbf{J}$, and the growth rates follow the eigenvalue moduli instead.
For an affine rule the Lyapunov spectrum is thus not an asymptotic
quantity to be approximated but a fixed property of one matrix: no trajectory is
simulated, no cut-off is chosen, and the numerical instability of the Jacobian
product never arises.

By Eq.~\eqref{eq:exact-spectrum}, the sum of the exponents equals $\ln\lvert\det\mathbf{J}\rvert$, the rate at which a volume of perturbations grows in tangent space. For a smooth map, tangent space is the local picture of phase space, so this is also the rate at which a small region of phase space grows or shrinks. For a Boolean map the two differ. In configuration space the analogue of volume is the number of configurations, which is preserved when no two configurations are mapped onto the same one, that is, when $\Phi$ is a bijection and the rule is reversible. For an affine rule the difference pattern evolves exactly as $\Delta\mathbf{s}^{t+1}=\mathbf{J}\,\Delta\mathbf{s}^{t}$ over $\mathrm{GF}(2)$, so two configurations merge when their nonzero difference pattern vanishes in one step. Such a pattern exists if and only if $\mathbf{J}$ is singular modulo two, which happens when $\det\mathbf{J}$ is even. Over the integers, $\mathbf{J}$ therefore sets the growth of tangent-space volume, and modulo two it decides whether the rule is reversible. Rule $150$ with $N$ coprime to three, for example, is reversible, yet its tangent-space volume triples at every step.

\subsection{Closed-form Lyapunov spectrum of affine ECAs}

With periodic boundary conditions, the constant Jacobian of the affine ECAs defined in Eq.~\eqref{eq:affine-eca} is the \emph{circulant matrix} in which every row carries the gradient, shifted by one cell,
\begin{equation}
    \label{eq:circulant-J}
    \mathbf{J}=
    \begin{pmatrix}
    a_\circ     & a_+       & 0         & \cdots    & 0         & a_-\\
    a_-         & a_\circ   & a_+       & 0         & \cdots    & 0\\
    0           & a_-       & a_\circ   & a_+       &           & \vdots\\
    \vdots      &           & \ddots    & \ddots    & \ddots    & 0\\
    0           &           &           & a_-       & a_\circ   & a_+\\
    a_+         & 0         & \cdots    & 0         & a_-       & a_\circ
    \end{pmatrix}.
\end{equation}
A circulant matrix is normal, so its singular values are the absolute values of its eigenvalues, and the eigenvalues of a circulant matrix are the DFT of its first row \cite{davis1979circulant}. Writing the eigenvalue at index $k$ as the gradient evaluated at the spatial frequency $k/N$,
\begin{equation}
    \label{eq:eca-exp}
    \lambda_k=a_\circ + a_+\exp\left(\frac{2\pi i k}{N}\right) + a_- \exp\left(\frac{-2\pi i k}{N}\right),
\end{equation}
and taking the modulus gives the singular values
\begin{equation}
    \label{eq:eca-singular-values}
    \begin{split}
    \sigma_k=\Biggl[\left(a_\circ
    +\bigl[a_+ + a_-\bigr]
    \cos\!\left(\frac{2\pi k}{N}\right)\right)^{2}\\
    \,+\,\left[a_+ - a_-\right]^{2}
    \sin^{2}\!\left(\frac{2\pi k}{N}\right)\Biggr]^{1/2},
    \end{split}
\end{equation}
for $k=1,\dots,N$, with the corresponding Lyapunov spectrum $\Lambda_k=\ln(\sigma_k)$. As $N\to\infty$ the variable $k/N$ becomes the continuous unit interval, and the spectrum depends on the rule only through its gradient.

\begin{figure}[t]
    \centering
    \includegraphics[width=\linewidth]{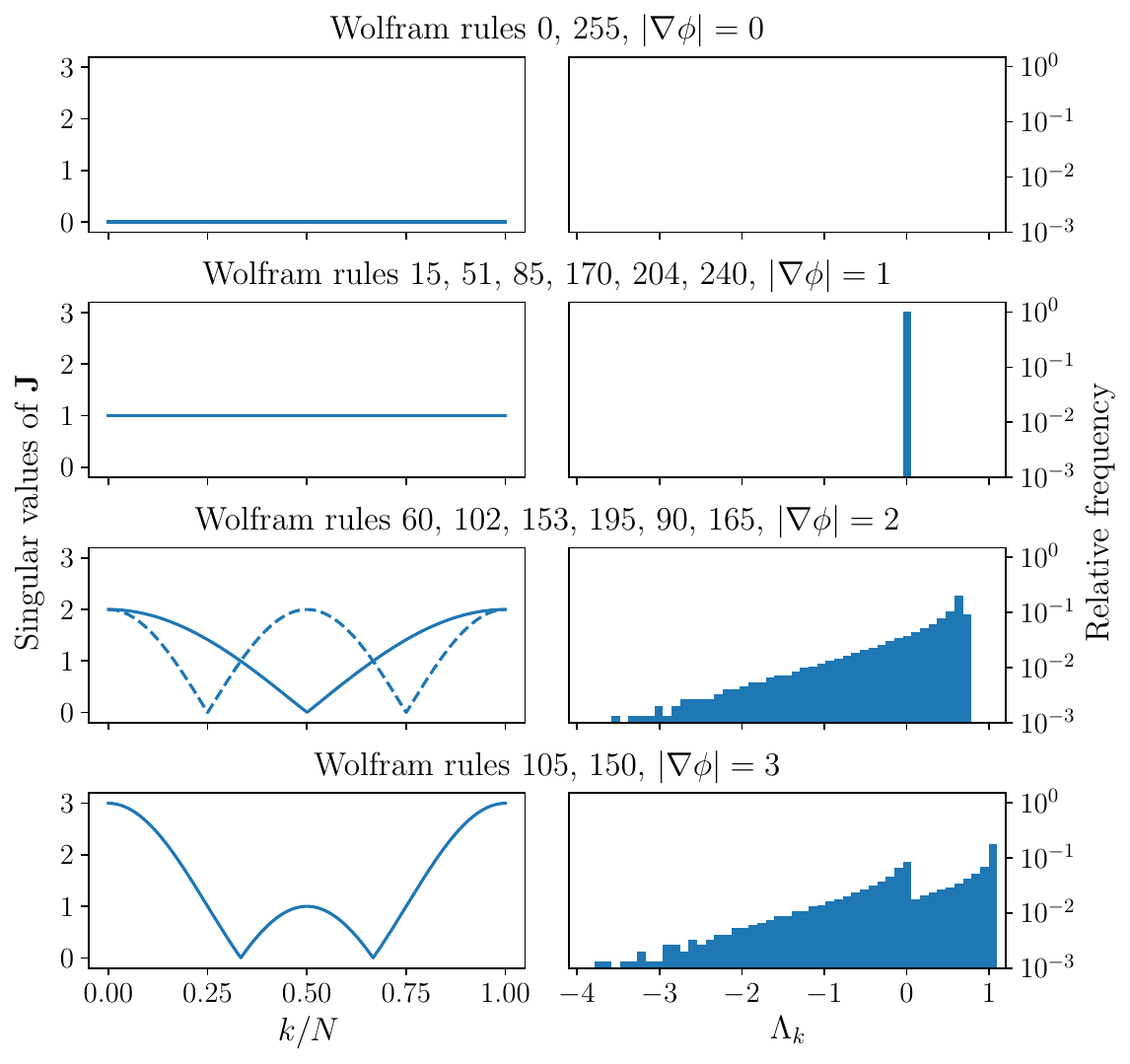}
    \caption{Singular values of the constant Jacobian of the affine ECAs
    (Tab.~\ref{tab:affine-ecas}) and their Lyapunov spectra, ordered by gradient weight, for $N=3001$. Rules with $|\nabla\phi|=2$ have two distinct singular value parametrisations  (full and dashed line), with identical associated spectra as $N\to\infty$.}
    \label{fig:eca-spectra}
\end{figure}

The \emph{gradient weight} $\lvert\nabla\phi\rvert$ organises the $16$ spectra into four distinct classes, as we show in Fig.~\ref{fig:eca-spectra}. It also fixes the maximal exponent: at $k=N$ every term in Eq.~\eqref{eq:eca-exp} is in phase, so $\sigma_N=\lvert\nabla\phi\rvert$ and $\mathrm{MLE}=\ln\lvert\nabla\phi\rvert$. The self-inclusive parity rule $150$ and its affine complement $105$, both with gradient $(1,1,1)$, are the most sensitive cases. Here Eq.~\eqref{eq:eca-singular-values} reduces to $\sigma_k=\lvert 1+2\cos(2\pi k/N)\rvert$, so
\begin{align}
    \label{eq:closed-form-150}
    \Lambda_k=\ln\left\lvert 1+2\cos\left(\frac{2\pi k}{N}\right)\right\rvert
\end{align}
and the $\mathrm{MLE}=\ln(3)$.

Singular values can vanish, and we take $\ln 0=-\infty$. For rule $150$, for instance, $\sigma_k$ vanishes when $N$ is a multiple of three; the sizes used below ($N=3001$ and $N=101$) avoid this, so every plotted exponent is finite.

\begin{figure}[t]
    \centering
    \includegraphics[width=\linewidth]{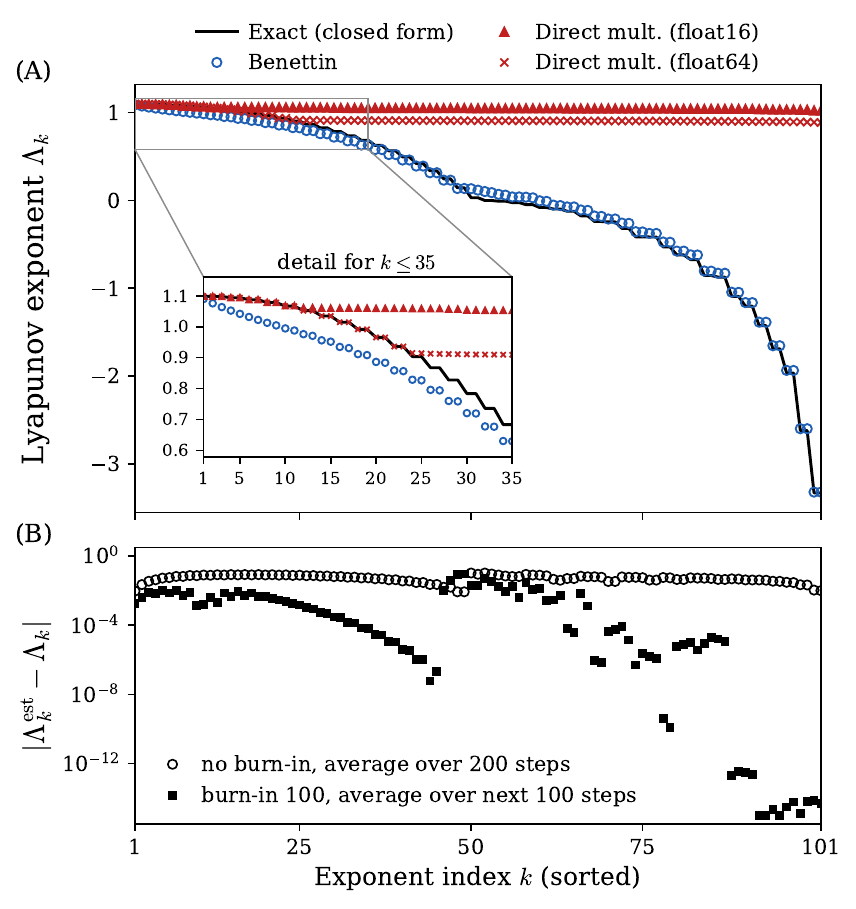}
    \caption{(A) Lyapunov spectrum of rule $150$ ($N=101$, sorted descending), computed at a common horizon $T=200$ by three methods: the exact closed form from Eq.~\eqref{eq:closed-form-150}, direct multiplication at two floating-point precisions, and Benettin's algorithm without burn-in. (B) Absolute error between the exact Lyapunov exponents and those estimated using Benettin's algorithm with or without a burn-in period of $100$ timesteps.}
    \label{fig:benchmark}
\end{figure}

The affine spectra give a controlled test of any numerical Lyapunov routine, because the exact answer is known. Fig.~\ref{fig:benchmark} compares, for rule $150$ ($N=101, T=200$), the closed-form spectrum with direct-multiplication approximations for 16-bit and 64-bit precision. As is well known, direct multiplication yields correct results for large exponents, but for small exponents the values plateau as the slower-growing directions are lost to floating-point error. Comparison with our analytical results also revealed errors in the spectra reported in a recent study \cite{vispoel2024damage}, which we correct in App.~\ref{app:vispoel-correction}.

Benettin's algorithm \cite{benettin1980lyapunov1,benettin1980lyapunov2} estimates the growth rates of the singular values of a long matrix product by re-orthonormalising the frame at regular intervals (here at every step), which avoids the two failure modes of direct multiplication: floating-point overflow, and the collapse of every column onto the dominant direction. It is an estimator with a transient: from an arbitrary starting frame the running average carries a bias until the frame has aligned, which a burn-in removes. The remaining error is governed by the ratio of consecutive singular values. Fig.~\ref{fig:benchmark}(A) shows values $\Lambda_k$ calculated with Benettin's algorithm, again for $T=200$ timesteps and without a burn-in. In Fig.~\ref{fig:benchmark}(B) we show the absolute error of the estimation compared to the analytical result, both with and without a burn-in of $100$ timesteps; without a burn-in, the initial bias makes the error larger. The sum of the exponents is a further simple check. For rule $150$ with $N=101$ it equals $\ln(3)\approx1.0986$; Benettin's algorithm preserves this value by construction, whereas direct multiplication in double precision at $T=200$ gives about $95$.

We stress that in reality one would not apply the algorithm to a product of identical matrices $\mathbf{J}$, as a single eigendecomposition yields $\ln|\lambda_k|=\ln\sigma_k$ directly, and the algorithm is unnecessary rather than merely fast. Additionally, for non-identical matrices one would use a larger number of timesteps and a larger burn-in time. For the sake of comparison, with a burn-in of $B=1000$ steps followed by averaging over $1000$ steps, Benettin's algorithm recovers the full spectrum of rule $150$ at $N=101$ to a maximum absolute error of $6.5 \times 10^{-5}$ across all $101$ exponents. The value of running it here is as a \textit{benchmark} that helps to make informed decisions on parameters of algorithms and allows critical review of their output, as in App.~\ref{app:vispoel-correction}.

\subsection{Generalisation to CAs with higher dimensions and arbitrary neighbourhoods}

The construction extends without difficulty to larger sizes of shift-invariant neighbourhoods in lattices of arbitrary dimensions. A neighbourhood $\mathcal{N}$ is specified by a set of offset vectors $\boldsymbol{\delta}=(\delta_1,\dots,\delta_D)\in\mathbb{Z}^D$, one per cell, including the central cell ($\boldsymbol{\delta}=\mathbf{0}$), where $\delta_d$ is the displacement along the $d$-th dimension; for the ECA the offsets are $\delta_1 \in \{-1, 0, +1\}$, which in Sec.~\ref{sec:jacobian} we referred to as cells $c_-, c_\circ$, and $c_+$. If we order the cells of a $D$-dimensional periodic lattice with side lengths $N_1,\dots,N_D$ lexicographically, so the lattice has $N=\prod_{d=1}^{D}N_d$ cells, an affine rule's constant Jacobian is then a $D$-level circulant matrix: a circulant matrix whose blocks are themselves circulant, $D$ times over. Such a matrix is diagonalised by the $D$-dimensional DFT \cite{davis1979circulant}, indexed by a frequency vector $\mathbf{k}=(k_1,\dots,k_D)$ with $k_d=1,\dots,N_d$, and its singular values are
\begin{equation}
    \label{eq:general-d}
    \sigma_{\mathbf{k}}
    =\Biggl\lvert\sum_{\boldsymbol{\delta}}a_{\boldsymbol{\delta}}\,
    \exp\Bigl(2\pi i\sum_{d=1}^{D}\frac{\delta_d k_d}{N_d}\Bigr)\Biggr\rvert,
\end{equation}
where the sum goes over the neighbourhood offsets $\boldsymbol{\delta}$ and $a_{\boldsymbol{\delta}} \in \{0,1\}$ is the sensitivity of the output to the cell at offset $\boldsymbol{\delta}$. Eq.~\eqref{eq:general-d} is the modulus of the multidimensional DFT of the gradient stencil; it is the
convolution theorem applied to the shift-invariant structure of the rule.

The maximal exponent is again attained where every exponential is in phase, at
$\mathbf{k}=(N_1,\dots,N_D)$, so
\begin{equation}
    \label{eq:mle-count}
    \mathrm{MLE}=\ln\Bigl(\sum_{\boldsymbol{\delta}}
    a_{\boldsymbol{\delta}}\Bigr).
\end{equation}
This is simply the logarithm of the number of inputs to which the rule is sensitive. For a self-inclusive parity rule every cell in the neighbourhood is a sensitive input, so $\mathrm{MLE}=\ln\lvert\mathcal{N}\rvert$, the logarithm of the neighbourhood size. As an illustration, the parity rule with the 2-D Moore neighbourhood has $\mathrm{MLE}=\ln(9)=2\ln(3)$, twice the value for ECA rule $150$. This reflects the fact that the Moore neighbourhood is the Cartesian product of two 1-D radius-one neighbourhoods; the MLE is additive over such a product.

\subsection{Bridging case: affine outer-totalistic rules}
\label{subsec:bridging-case}

The rule space of CAs with extended neighbourhoods grows rapidly with $|\mathcal{N}|$, quickly becoming too large to study exhaustively. Most of these rules are also anisotropic, and so of limited interest to applications where the orientation of a neighbour carries no physical meaning. For both reasons, such CAs are commonly studied in \emph{outer-totalistic} form, where the update depends only on the central state and on the \emph{sum} of the surrounding states \cite{packard1985two}. An affine rule is outer-totalistic when it is equally (in)sensitive to every neighbour, that is, when all the directional gradient entries share a common value $g\in\{0,1\}$. Its sensitivity to the central cell, $a_\circ\in\{0,1\}$, is a separate choice: $a_\circ=1$ if the output depends on the cell's own state and $a_\circ=0$ if it does not. Every affine outer-totalistic rule is then fixed, up to the constant $a_0$, by the pair $(a_\circ,g)$, independently of which neighbourhood is used.

\begin{figure}[t]
\centering
    \includegraphics[width=\linewidth]{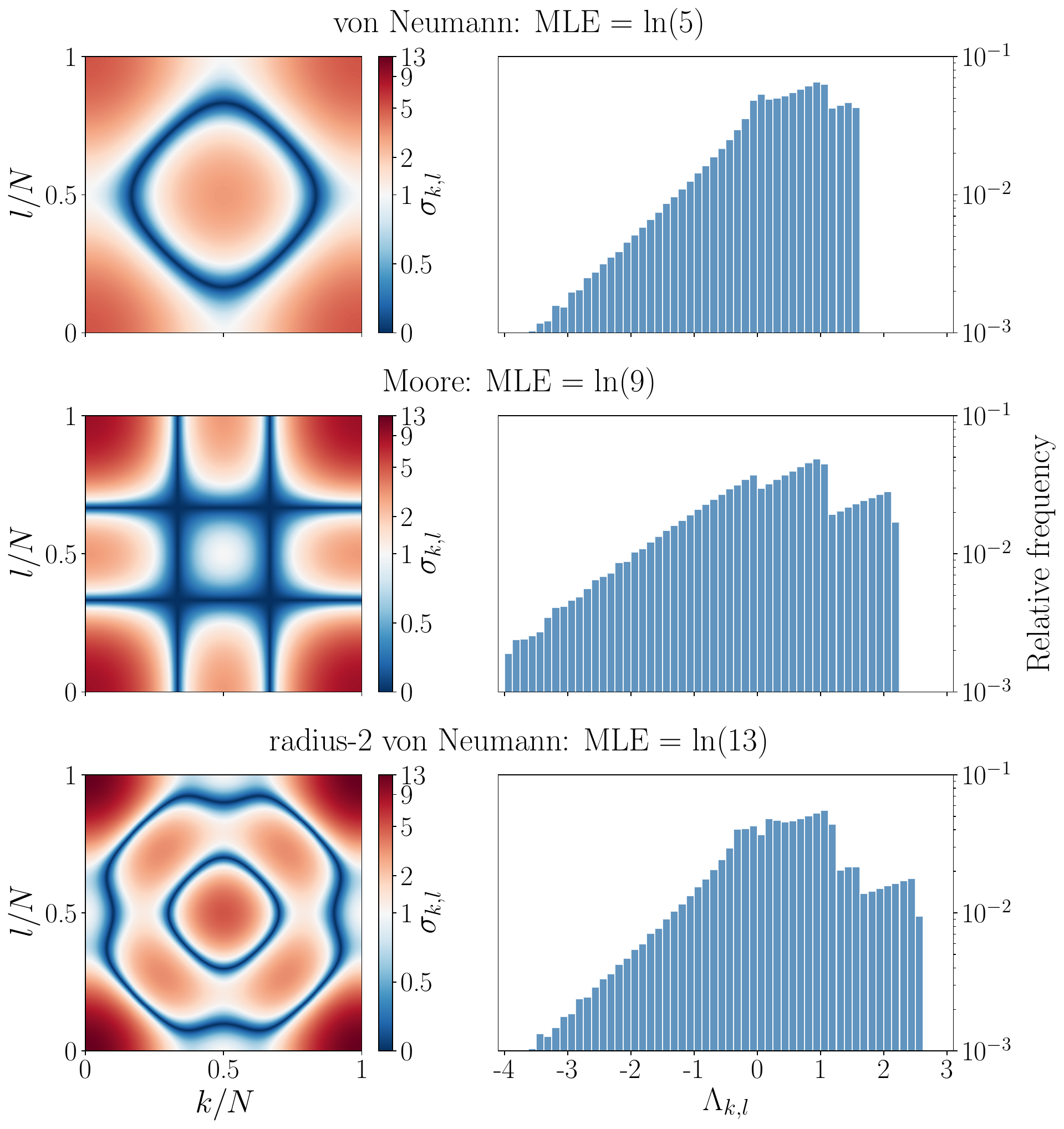}
    \caption{Singular values and Lyapunov spectra of the self-inclusive parity
    rule on 2-D lattices, for the three neighbourhoods of
    Tab.~\ref{tab:structure-factors}, with $N=201$ cells per side. In the Moore case $800$ of the $40\,401$ singular values vanish, since $3$ divides $N$, and their exponents of $-\infty$ are not shown.}
    \label{fig:2d-parity}
\end{figure}

Consider three standard neighbourhoods in 2-D CAs on a square lattice of side $N$ (so $N_1=N_2=N$): von Neumann, Moore, and radius-2 von Neumann, defined through respectively 5, 9, and 13 offset vectors $\boldsymbol{\delta}$. Because these are symmetric under $\boldsymbol{\delta}\mapsto-\boldsymbol{\delta}$, the exponentials in Eq.~\eqref{eq:general-d} pair into cosines, and the singular values become
\begin{equation*}
    \label{eq:ot-singular}
    \sigma_{k,l}=\bigl\lvert a_\circ+g\,\mathcal{K}(k,l)\bigr\rvert,\ \ \ 
    \mathcal{K}(k,l)=\sum_{\boldsymbol{\delta}\neq\boldsymbol{0}}
    \cos\left(\frac{2\pi(\delta_1 k+\delta_2 l)}{N}\right),
\end{equation*}
where $\mathcal{K}$ is the DFT of the indicator function of the neighbourhood without its central cell (the neighbourhood's \emph{structure factor}), and we wrote $(k_1, k_2) = (k,l)$. The structure factor is a property of the neighbourhood alone; the rule enters only through the pair $(a_\circ,g)$. The parity rule is the case $g=1$: the self-inclusive parity rule has $a_\circ=1$ and is totalistic (it counts the central cell in the sum), whilst the self-exclusive parity rule has $a_\circ=0$ and is genuinely outer-totalistic (blind to the central cell). Tab.~\ref{tab:structure-factors} gives $\mathcal{K}$ and the self-inclusive parity-rule MLE for the three neighbourhoods, whilst Fig.~\ref{fig:2d-parity} shows the corresponding spectra. The fact that these rules are both (outer)totalistic and affine allows for the analytical study of the Lyapunov spectra of parity rules on arbitrary graphs.

\begin{table}[t]
    \centering
    \caption{Structure factor $\mathcal{K}(k,l)$ and self-inclusive parity-rule
    maximal Lyapunov exponent for three neighbourhoods, with $\alpha=2\pi k/N$
    and $\beta=2\pi l/N$.}
    \label{tab:structure-factors}
    \begin{tabular*}{\tblwidth}{@{} l l r @{}}
    \toprule
    $\mathcal{N}$ & $\mathcal{K}(k,l)$ & \textbf{Parity MLE}\\
    \midrule
    von Neumann      & $2\cos\alpha+2\cos\beta$ & $\ln(5)$\\[2pt]
    Moore            & $2\cos\alpha+2\cos\beta+4\cos\alpha\cos\beta$ & $\ln(9)$\\[2pt]
    radius-2 v.\ N.  & $\begin{aligned}[t]&2\cos\alpha+2\cos\beta+4\cos\alpha\cos\beta\\
                       &\quad+2\cos 2\alpha+2\cos 2\beta\end{aligned}$ & $\ln(13)$\\
    \bottomrule
    \end{tabular*}
\end{table}

% ============================================================
\section{The parity rule on networks}
\label{sec:parity-network}

\subsection{The parity rule}

Consider a simple, undirected, unweighted graph with $N$ nodes and adjacency matrix $\mathbf{A}\in\{0,1\}^{N\times N}$. The parity rule updates a node according to the parity of the state sum over its neighbourhood,
\begin{equation}
    \label{eq:parity-rule}
    s_i^{t+1}=a_0\oplus\, a_\circ s_i^t\,\oplus\!\!\bigoplus_{j:\,v_j\in\mathcal{N}(v_i)}\!\!s_j^t,
    \qquad a_0, a_\circ\in\{0,1\}.
\end{equation}
Here $\mathcal{N}(v_i)$ is the set of nodes adjacent to $v_i$ so, unlike a CA neighbourhood, it does not contain the node itself. When $a_\circ = 1$ the rule is self-inclusive (totalistic); otherwise it is self-exclusive (outer-totalistic). If the sum of the states of all connected nodes is even, the rule maps state $s_i^t$ to $a_0\oplus a_\circ s_i^t$, and to $a_0\oplus a_\circ s_i^t \oplus 1$ otherwise. As in Eq.~\eqref{eq:affine-eca}, the constant $a_0$ does not enter the Boolean derivative, so the two complementary cases share the same Jacobian, and hence the same perturbation dynamics; they differ only by a state-independent background that the derivative removes. Note that for a ring graph of degree $2$, the case $a_0 = a_\circ = 0$ corresponds exactly to ECA rule $90$ and the case $a_0 = 0, a_\circ = 1$ to rule $150$, illustrating that indeed Eq.~\eqref{eq:parity-rule} is the network generalisation of the parity CAs. This equivalence can be generalised to the parity rules ($g=1$) on symmetric neighbourhoods in $D$ dimensions from Subsec.~\ref{subsec:bridging-case} by choosing as underlying topology the Cayley graph of $(\mathbb{Z}/N\mathbb{Z})^D$ with the neighbourhood, without its central cell, as connection set.

By the argument of Subsec.~\ref{sec:affine} the Jacobian associated with rule \eqref{eq:parity-rule} is constant. Evaluating Eq.~\eqref{eq:boolean-derivative} on Eq.~\eqref{eq:parity-rule} shows that this constant Jacobian is the adjacency matrix, plus the identity in the self-inclusive case, $\mathbf{J}=\mathbf{A} + a_\circ\mathbf{I}$: flipping any neighbour changes the parity of the (outer)totalistic neighbourhood sum and therefore toggles the output, whilst flipping a non-neighbour does not directly affect the node's state.

\subsection{The Lyapunov spectrum and the absolute graph spectrum}

With $\mathbf{J}=\mathbf{A} + a_\circ\mathbf{I}$, the exact spectrum \eqref{eq:exact-spectrum} follows directly. The matrix $\mathbf{A}+a_\circ\mathbf{I}$ is real and symmetric, with eigenvalues $\lambda_k(\mathbf{A})+a_\circ$, so its singular values are their absolute values, and
\begin{equation}
    \label{eq:network-spectrum}
    \Lambda_k=\ln\bigl(\sigma_k(\mathbf{A} + a_\circ\mathbf{I})\bigr)
    =\ln\bigl\lvert\lambda_k(\mathbf{A})+ a_\circ\bigr\rvert
\end{equation}
for $k=1,\dots,N$. The Lyapunov spectrum of a parity rule on a graph is, up to a logarithm, an absolute value and the shift $a_\circ$, the spectrum of that graph \cite{vanmieghem2011graph}. The maximal exponent is $\mathrm{MLE}=\ln\rho(\mathbf{A}+a_\circ\mathbf{I})$, where the spectral radius of the shifted matrix is $\rho(\mathbf{A}+a_\circ\mathbf{I})=\max_k\lvert\lambda_k(\mathbf{A})+a_\circ\rvert=\lambda_N+a_\circ$, since $\mathbf{A}$ is nonnegative with largest eigenvalue $\lambda_N=\rho(\mathbf{A})$. The spectral radius of any graph is at least its mean degree \cite{collatz1957spektren}, so the parity rule has a positive MLE on every graph with mean degree above one. Note again that the finite system in configuration space is eventually periodic, so the MLE expresses exponential sensitivity in tangent space, not chaos in the topological sense. Our contribution here is not a diagnostic for actual dynamical processes on networks, but rather an edge-case mathematical connection between distinct concepts: the Lyapunov spectrum for dynamical processes on the one hand, and spectral theory related to walk counting in network science on the other.

\section{Conclusion}
\label{sec:conclusion}

We have shown that the affine Boolean rules, those whose update is an \textsc{xor} of inputs and a constant, have a configuration-independent Boolean Jacobian, and therefore a Lyapunov spectrum that is exactly the logarithms of the eigenvalue moduli of one constant matrix. For an affine CA on a periodic lattice that matrix is multilevel circulant, so the spectrum is the log-modulus of the multidimensional DFT of the gradient stencil. For the parity rule on an arbitrary graph it is the adjacency matrix, plus the identity in the self-inclusive case, and the Lyapunov spectrum is the logarithm of the absolute spectrum of that matrix. In other words, affine rules are the analytically tractable corner of rule space, and the value of these exact results is that they fix reference points where Lyapunov spectra are otherwise only approximated, having shown their merit by exposing artefacts in direct multiplication and a transient bias in Benettin's algorithm.

Three new directions follow naturally. First, the clean affine case for CAs is a starting point for rules that are affine only on part of their neighbourhood, or only along the trajectory actually visited; there the affine spectrum may act as a leading-order term to be reconciled with a numerical estimate. Second, results may be further generalised to directed graphs or non-deterministic systems, for which the Jacobian may be non-normal or contain nonbinary elements. Note, for example, that the limit in Eq.~\eqref{eq:lyapunov-spectrum} also holds for non-normal matrices due to Yamamoto's theorem \cite{yamamoto1967extreme}. Third, exponents have been used to classify CAs \cite{vispoel2022progress}, and as objects that sit somewhere between a genotype and a phenotype parameter, Lyapunov spectra may be an interesting tool in the classification of CAs and other DDSs. The benchmark results in this work will, in any case, serve as a self-contained sanity check.

% ============================================================
\section*{Declaration of competing interest}
The authors declare that they have no known competing financial interests or personal relationships that could have appeared to influence the work reported in this paper.

\section*{Data availability}
The Python code used to compute the Lyapunov spectra and to reproduce the figures in this article is available through the GitHub repository mrollier/exact-lyapunov-spectra.

\section*{Acknowledgements}
This work has been partially supported by the FWO grant with project title ``An analysis of network automata as models for biological and natural processes'' [3G0G0122]; by the FWO travel grant with file name V412625N; by the FWO congress participation grant K105625N. The authors wish to thank Dr.~Milan Vispoel for the elucidating informal discussions that laid the groundwork for the present analysis.

\appendix

\section{Correction of misprints in Vichniac (1990)}
\label{app:vichniac}

Tab.~1 in Vichniac's seminal work \cite{vichniac1990boolean} contains seven incorrect gradient entries, spread over the five ECA rules $62$, $110$, $130$, $146$ and $172$. Here we derive the correct entries using Vichniac's fifth method and indicate how his third method (using symmetries) probably resulted in most of the errors.

Rule $62$ corresponds to the binary string $00111110$ in the Wolfram convention. Calculating $\partial \Phi_i / \partial s_{i-1}$ corresponds to inspecting what happens to the output of the local update rule when the left-most cell state is complemented. This is equal to bitwise \textsc{xor}ing the first half of the binary string with the second: $(0011) \oplus (1110) = (1101)$, such that we want to find a function that maps $(1,1)\mapsto 1$, $(1,0)\mapsto 1$, $(0,1)\mapsto 0$, and $(0,0)\mapsto 1$. With the notational convention of both Vichniac's Tab.~1 and our Tab.~\ref{tab:gradient-table} (writing $x$ for $s$), this function takes the general form $(x_i, x_{i+1}) \mapsto x_i + \bar{x}_{i+1}$. The overbar in Vichniac's entry is on the wrong variable, as he likely used his own third method from rule $56$ but skipped a substitution.

Rule $110$ corresponds to the binary string $01101110$ and its printed gradient contains two misprinted entries. Applying the same technique as before, we find for $\partial \Phi_i / \partial s_{i-1}$ the entry $x_i x_{i+1}$, as opposed to the original entry $\bar{x}_i x_{i+1}$. To find $\partial \Phi_i / \partial s_{i+1}$ we need to inspect what happens to the local update rule's output when we complement the right-most cell state. This corresponds to bitwise \textsc{xor}ing all odd-position binary string bits with the subsequent even-position bit: $(0111) \oplus (1010) = (1101)$, requiring a function that maps $(1,1)\mapsto 1$, $(1,0)\mapsto 1$, $(0,1)\mapsto 0$, and $(0,0)\mapsto 1$. This results in the entry $x_{i-1} + \bar{x}_i$, which is different from Vichniac's $x_{i-1} + x_i$. Both errors arguably result from wrongly reflecting the solution for rule $44$.

Rule $130$ corresponds to the binary string $10000010$, and for $\partial \Phi_i / \partial s_{i+1}$ we find the entry $x_{i-1}x_i + \bar{x}_{i-1}\bar{x}_i$. Comparing with Vichniac's entry, this is probably just a subscript typo.

Rule $146$ corresponds to the binary string $10010010$. For $\partial \Phi_i / \partial s_{i-1}$ we find $\bar{x}_i + x_{i+1}$ instead of Vichniac's $x_i + x_{i+1}$, and for $\partial \Phi_i / \partial s_{i+1}$ we find $x_{i-1} + \bar{x}_i$ instead of $x_{i-1} + x_i$. The original entries were probably obtained from rule $104$ by method 3 without the appropriate complementation.

Rule $172$ corresponds to the binary string $10101100$. For $\partial \Phi_i / \partial s_{i-1}$ we find $x_i \bar{x}_{i+1} + \bar{x}_i x_{i+1}$ instead of Vichniac's $x_i x_{i+1} + \bar{x}_i \bar{x}_{i+1}$, probably resulting from reflecting the result from rule $78$ but forgetting the complementation of the central cell state.

We invite the reader to verify these results with the code provided in our associated GitHub repository: running \verb|python verify_vichniac.py| recomputes all $88$ gradients from the definition and lists the seven entries that differ from the published table, and the test suite (\verb|pytest|) reproduces the additivity and symmetry cross-checks.

\section{Perturbations in configuration vs.~tangent space}
\label{app:mle-vs-damage}

\begin{figure}
    \centering
    \includegraphics[width=\linewidth]{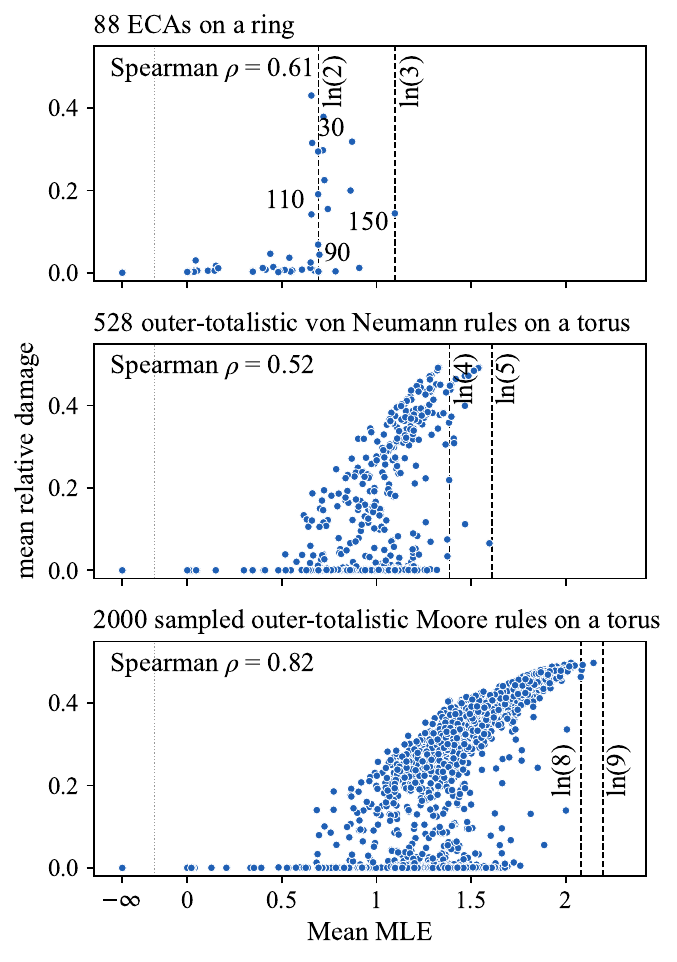}
    \caption{The mean relative damage (fraction of defect cells in the light cone) in configuration space versus the mean MLE in tangent space, for three CA types. The vertical dashed lines indicate the theoretical maxima for the self-inclusive or self-exclusive parity rule.}
    \label{fig:mle-vs-damage}
\end{figure}

The defect cone in configuration space and the MLE calculated in tangent space are separate entities but do correlate, as Fig.~\ref{fig:mle-vs-damage} shows for ECAs and 2-D outer-totalistic CAs. For every rule the mean relative damage caused by flipping one cell (differing cells over the maximal light cone, averaged over the final steps and over random initial configurations) is plotted against the MLE estimated from a single tangent vector on the same trajectories, on lattices large enough that the damage never wraps: all $88$ non-equivalent ECAs ($N = 607$, $T = 300$), all $528$ non-equivalent outer-totalistic von Neumann rules ($L = 149$, $T = 70$), and $2000$ of the $131\,328$ non-equivalent outer-totalistic Moore rules, sampled uniformly since that family is too large to enumerate. Rules whose Jacobian annihilates the tangent vector have an MLE of $-\infty$ and occupy a separate column; dashed lines mark the exact exponents of the affine and parity rules, which combine a large MLE with comparatively little damage, their defect patterns being Sierpi\'nski-like. The correlation is positive but far from determinative (Spearman's $\rho = 0.61$, $0.52$ and $0.82$ respectively): at a given MLE the damage ranges from nil to nearly half the cone. Please refer to our GitHub repository for further details.

\section{Correction of spectra in Vispoel et al.~(2024)}
\label{app:vispoel-correction}

\begin{figure}
    \centering
    \includegraphics[width=\linewidth]{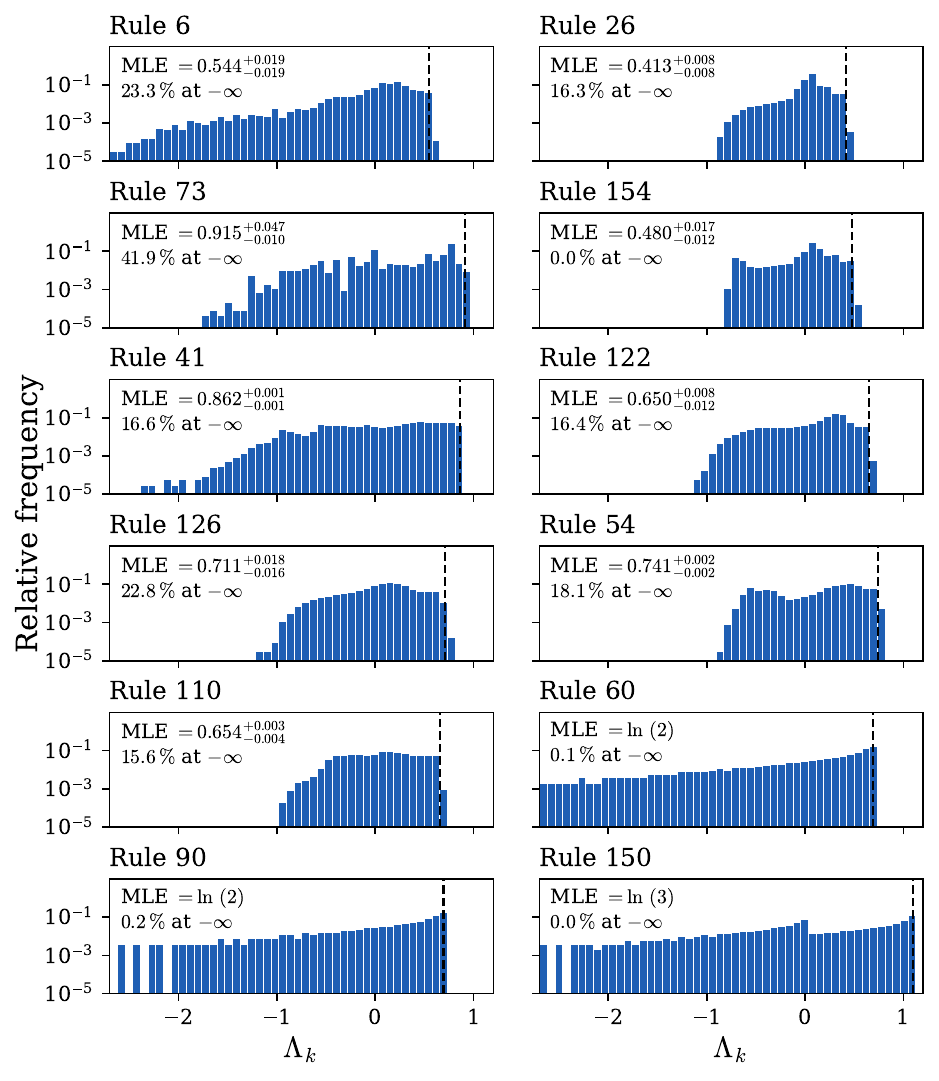}
    \caption{Corrected Lyapunov spectra of the rules selected by Vispoel et al.~\cite{vispoel2024damage}. Every panel indicates the MLE (dashed line) and the percentage of vanishing singular values; for the non-affine rules the MLE is a mean, with $1\sigma$-equivalent bounds.}
    \label{fig:lyapunov_spectra_nonaffine_ecas}
\end{figure}

Whilst Vispoel et al.'s research paper \cite{vispoel2024damage} provides an excellent and didactic introduction to the meaning of the Lyapunov exponent in the context of discrete dynamical systems, we discovered a number of incorrectly calculated spectra of the affine rules $60$, $90$ and $150$ through comparison with our analytical findings, and subsequently found that other spectra were also affected. In particular, the reported MLEs are incorrect, and the tail for low $\Lambda_k$ values is missing. We briefly correct these in Fig.~\ref{fig:lyapunov_spectra_nonaffine_ecas}, with rules $60$, $90$ and $150$ from their exact spectra and the others using Benettin's algorithm, and use the same rules and parameter values as in the original work: $N=1000$, $T=500$, averaging over realisations from $40$ random initial configurations, for rules $6$, $26$, $73$ and $154$ (Class II), rules $41$, $122$ and $126$ (Class III), and rules $54$ and $110$ (Class IV). Unlike the original work, we discard the first $200$ timesteps as a burn-in. Spectra for all $88$ non-equivalent ECAs can be found and re-calculated in the repository as well.

\printcredits

\bibliographystyle{cas-model2-names}
\bibliography{lyapunov_R2}

\end{document}